# Picosecond operations on superconducting quantum register based on Ramsey patterns

*M.V. Bastrakova[1], N.V. Klenov[2-5], V.I. Ruzhickiy[2,4], I.I. Soloviev[3,4,*], and A.M. Satanin[4,6]*

[1] Lobachevsky State University of Nizhny Novgorod, Nizhny Novgorod 603950, Russia
[2] Faculty of Physics, Lomonosov Moscow State University, Moscow 119991, Russia
[3] Lomonosov Moscow State University Skobeltsyn Institute of Nuclear Physics, Moscow 119991, Russia
[4] Dukhov All-Russia Research Institute of Automatics, Moscow 101000, Russia
[5] Moscow Technical University of Communication and Informatics, 111024 Moscow Russia
[6] Russia National Research University Higher School of Economics, 101000 Moscow, Russia

An ultrafast qubit control concept is proposed to reduce the duration of operations with a single and multiple superconducting qubits. It is based on the generation of Ramsey fringes due to unipolar picosecond control pulses. The key role in the concept is played by the interference of waves of qubit states population propagating forward and backward in time. The influence of the shape and duration of control pulses on the contrast of the interference pattern is revealed in the frame of Ramsey's paradigm. Protocols for observation of Ramsey oscillations and implementation of various gate operations are developed. We also suggest a notional engineering solution for creating the required picosecond control pulses with desired shape and amplitude. It is demonstrated that this makes it possible to control the quantum states of the system with the fidelity of more than 99 %.

## I. INTRODUCTION

The basis of modern methods for controlling registers of superconducting quantum computers (QC) is the Rabi technique [1-6]: artificial quantum systems (with a set of characteristic transitions frequencies, $\omega_{ij}$, between the states $i$ and $j$), are affected by modulated pulses of an electromagnetic field, which carrier frequency, $\omega$, is close to the qubit one, $\omega_{ij}$. This approach has already allowed the development of algorithmic intermediate-scale QC containing up to 72 superconducting qubits [7-10]. Google team recently demonstrated quantum supremacy for the first time in solving a limited range of mathematical problems [11]. This has become possible since with the using of the Rabi technique the fidelity reaches 99.9 % for single-qubit gates, and 99 % for two-qubit gates.

However, the problems with logical operations in QC impede its possible practical applications. In particular, it is difficult to increase the number of control microwave channels come from room temperature that is necessary to increase the "depth" of the quantum circuit. In addition, in the framework of the Rabi technique, the duration of a logical operation is always much longer than the inverse qubit transition frequency, $\omega_{ij}^{-1}$. Consequently, the problem of accelerating single- and multiple-qubit gates is also relevant, despite the fact that the decoherence time for modern qubits has already reached hundreds of microseconds.

While superconducting qubits are commonly fabricated from aluminum and have the transition frequencies in GHz range, utilization of superconductors with larger superconducting energy gap, e.g., niobium, opens access to electronics with sub-THz characteristic frequencies [12]. Such energy-efficient Nb-based superconducting digital devices were already proposed for the development of classical control processor of QC located in the refrigerator near quantum processor chip for increasing the number of control channels [13-18]. In this approach the Rabi technique is modified in such a way that long sequences of short single-flux-quantum (SFQ), $\Phi_0 = h/2e$ (where h is the Planck constant and e is the electron charge), pulses with the duration, $\tau_{op} \ll \omega_{ij}^{-1}$, are used to control the qubit. They can be created in the same chip with qubits or in an auxiliary chip placed at 4.2 K temperature stage using circuits of rapid single-flux-quantum (RSFQ) logic [19].

In this paper, we develop the ultrafast qubit control, where the time of operation, $\tau_{op}$, is much less than the characteristic transition frequency, $\tau_{op} \ll \omega_{ij}^{-1}$, and lies in a picosecond timescale, typical for SFQ pulses. Some background of the proposed concept is laid in our earlier studies of the implementation of the simplest single-qubit operations and read-out procedure [20-25]. We have investigated that a fairly short unipolar pulse with a wide spectrum, carrying a large energy, will always transfer a two-level system to a new state. We have shown how to select the amplitude and duration of such a pulse for the implementation of a complete set of single-qubit operations [22]-[24]. But it was impossible to implement all the necessary operations even in the simplest register (with two connected qubits) in this way. It turned out that we need pairs (or triples) of pulses with controlled parameters to implement a set of two-qubit operations.

It is worth emphasizing that we propose a significant change in the approach of the articles [13, 14]. Instead of hundreds of identical Delta-like pulses, we suggest using several pulses with controlled durations and amplitudes. Instead of numerical optimization of the time intervals between individual identical pulses, we use the analytical theory for Ramsey interference to find on/off moments and amplitudes for a small number of relatively long pulses.

The paper is organized as follows. Firstly, we present an analogy between the effect of modulated high-frequency control pulses and unmodulated rectangular unipolar control pulses acting on a single qubit.

Secondly, we suggest an analytical rationale for replacing the Rabi technique with the interference (the Ramsey) method. The mentioned Ramsey interference is considered in the frame of the following interpretation: the interference of two waves (describing the population of the qubit states) arises due to effect of a single unipolar pulse; one of the waves propagates forward and the other backward in time. We present analytical and numerical calculations demonstrating that the Ramsey fringes allowing implementation of a complete set of single-qubit operations can be realized by unipolar pulses.

Then we consider the Ramsey method in the context of fast control of the states of a simple two-qubit register. In this case, the formation of the Ramsey fringes is more complex, because the system evolution results from the interference of four waves with different frequencies. Since the frequencies of the waves depend on the coupling constant of qubits, Ramsey fringes also turns out to be dependent on the interaction of qubits.

Finally, we propose a notional engineering solution for creating unipolar pulses from SFQ pulses. The obtained magnetic flux pulses have close to rectangular shape, and their amplitude and duration can be independently adjusted for implementation of the considered operations with qubits.

## II. SINGLE-QUBIT OPERATIONS

### A. The analogy between the effect of modulated high-frequency and unmodulated unipolar pulses

If initially a quantum system is prepared in the state, $\left|\psi(0)\right\rangle$, then its temporal evolution is described by the equation:

$$i\frac{\partial\left|\psi(t)\right\rangle}{\partial t} = H(t)\left|\psi(t)\right\rangle, \; \hbar = 1, \quad (1)$$

where $H(t)$ is its time-dependent Hamiltonian. In general, the solution of Eq. (1) has the form

$$\left|\psi(t)\right\rangle = U(t)\left|\psi(0)\right\rangle, \; U(t) = \hat{P}\exp\left(-i\int_0^t dt_1 H(t_1)\right), \quad (2)$$

where $\hat{P}$ is the chronological ordering operator in the evolution operator, $U(t)$. If the system is affected by a sequence of control pulses interrupted by the free evolution, then the calculation of unitary rotation of the state vector for a full period of time is reduced to multiplying the evolution operators. We will use the general approach developed in [24] to calculate the evolution operator of a single- and multiple-qubit system.

The flux qubit Hamiltonian in an external field has the form [2, 3]:

$$H(t) = -\frac{1}{2}\left(\Delta\sigma_z + \varepsilon(t)\sigma_x\right), \quad (3)$$

where $\varepsilon(t)$ – corresponds to the external field, $\Delta$ – corresponds to the transition frequency, $\omega_{01}$, between basic states, $\left|0\right\rangle \equiv \left|\downarrow\right\rangle = \begin{pmatrix}1\\0\end{pmatrix}$ and $\left|1\right\rangle \equiv \left|\uparrow\right\rangle = \begin{pmatrix}0\\1\end{pmatrix}$, in the absence of disturbance; $\sigma_x$, $\sigma_z$ – Pauli matrices.

Let the "Rabi pulse" acts on the system:

$$\varepsilon(t) = Af(t)\cos(\omega t), \; f(0) = 1,$$

where $A$ is the characteristic amplitude.

It is well known, that if a resonant periodic weak field acts on a qubit (Rabi technique), one can remove all the fast rotating terms with a frequency and write an equation for slow amplitudes (the rotating wave approximation (RWA). This can be done by using the canonical transformation,

$$S(t) = \exp\left(\frac{i\omega t \sigma_z}{2}\right), \; \bar{H}(t) = S^\dagger H(t)S - iS^\dagger\frac{\partial S}{\partial t}. \quad (4)$$

RWA then gives:

$$\bar{H}(t) = -\frac{1}{2}\left((\Delta - \omega)\sigma_z + \frac{A}{2}\sigma_x\right). \quad (5)$$

For simplicity, we consider the case where the pulse envelope has a rectangular shape with the width $\tau$. Then the evolution operator can be written as:

$$U_1 = e^{-iM_1}, M_1 = M_{1x}\sigma_x + M_{1z}\sigma_z, M_{1x} = -A\tau/4, M_{1z} = -(\Delta - \omega)\tau/2.$$

The eigenvalues of the $M$-operator and explicit expression of the evolution operator is as follows:

$$\mu_1 = -\mu_0 = \mu, \; \mu = \sqrt{M_{1x}^2 + M_{1z}^2} = \frac{\tau}{2}\Omega_R, \; \Omega_R = \sqrt{(\Delta - \omega)^2 + (A/2)^2},$$

$$U_1 = \cos\mu\,\mathrm{I} - \frac{i\sin\mu}{\mu}M_1, \; \mathrm{I} = \begin{pmatrix}1 & 0\\0 & 1\end{pmatrix}. \quad (6)$$

The probability of as a function of time with Rabi frequency, $\Omega_R$, is determined by the well-known expression [1]:

$$W_{|0\rangle\to|1\rangle} = \frac{A^2}{\Omega_R^2}\sin^2\left(\frac{\tau}{2}\Omega_R\right). \quad (7)$$

Let us calculate the evolution operator for a single unmodulated rectangular pulse (unipolar pulse) with the duration $\tau$:

$$\varepsilon(t) = Af(t), \; f(t) = \Theta(t)\Theta(\tau - t),$$

where $\Theta(t)$ is the Heaviside step function. In this case by analogy with the Rabi pulse we have

$$U_1 = e^{-iR_1}, \; R = R_{1x}\sigma_x + R_{1z}\sigma_z, \; R_{1x} = -A\tau/2, \; R_{1z} = -\Delta\tau/2.$$

The eigenvalues of the $R$ – operator are

$$r_1 = -r_0 = r, \; r = \sqrt{R_{1x}^2 + R_{1z}^2} = \frac{\tau}{2}\Omega, \; \Omega = \sqrt{\Delta^2 + A^2},$$

$$U_1 = \cos r \cdot I - \frac{i\sin r}{r}R_1. \tag{8}$$

Now for the broadband impact, the role of the Rabi frequency, $\Omega$, is played by the amplitude of the pulse, $A$, while the role of the frequency detuning, $\Delta - \omega$, is played by the distance between the basic levels of the qubit, $\Delta$. The probability of transitions between the basic states in the case of a single unipolar pulse is determined by the expression:

$$W_{|0\rangle \to |\uparrow\rangle} = \frac{A^2}{\Omega^2}\sin^2\left(\frac{\tau}{2}\Omega\right). \tag{9}$$

Formally, the expression (9) coincides with the well-known Rabi formula [1]. Now for the short impact with broadband spectrum, the role of the Rabi frequency, $\Omega$, is played by the amplitude of the pulse, $A$, while the role of the frequency detuning is played by the distance between the basic levels of the qubit, $\Delta$. The main difference from the Rabi approach (RWA) is that the energy of the external field must be greater than the tunnel energy of the qubit, $A >> \Delta$, and the pulse duration must correspond to the frequency of the qubit as $\Delta\tau \ll 1$ [22, 23]. Naturally, the condition for the pulse amplitude from above is dictated only in order to prevent the qubit from going to the upper excited levels of the system.

In this paper, we mean that we consider flux qubits (3JJ qubits), which have a relatively large distance from the selected doublet to the highly excited levels [3]. At the same time, our calculations assumed that the amplitude of the external impact is significantly less than the energy between the excited qubit level and the other high-energy levels, which does not violate the two-level approximation.

At the moment it is appropriate to summarize the intermediate conclusions:
(i) The duration of logical operations under the Rabi technique is limited from below: the pulse duration must be much greater than the inverse carrier frequency of the pulse.
(ii) A two-level system can be controlled at picosecond times using unipolar pulses: a single pulse with sharp fronts can create a perfect contrast of populations in the plane of parameters.

### B. Ramsay interference: modulated high-frequency versus unmodulated unipolar pulses

The general nature of the phenomena underlying the Rabi and Ramsey interference can be seen by representing the qubit wave function as $|\psi(t)\rangle = \psi_0(t)|0\rangle + \psi_1(t)|1\rangle$. The amplitudes $\psi_0(t)$ and $\psi_1(t)$ obey the following equations:

$$i\frac{\partial\psi_0}{\partial t} = -\frac{1}{2}\big(\Delta\psi_0 + \varepsilon(t)\psi_1\big), \; i\frac{\partial\psi_1}{\partial t} = -\frac{1}{2}\big(-\Delta\psi_1 + \varepsilon(t)\psi_0\big). \tag{10}$$

For a single unipolar pulse, we can divide the domain of definition of $\psi_0(t)$ and $\psi_1(t)$ functions into three parts: where $\varepsilon(t)$ is a finite constant value ($0 < t < \tau$), and where it vanishes ($t < 0$, $t > \tau$). One can introduce the boundary conditions for amplitudes at the moments when the considered impact turns on and off. This makes the problem of this qubit state evolution similar to the problem of the spatial distribution of the wave function for the case of particle tunneling through a rectangular barrier. In particular, the equation for the amplitudes $\psi_1(t)$ and $\psi_0(t)$ at $0 < t < \tau$ can be written as:

$$\frac{\partial^2\psi_1}{\partial t^2} + k^2\psi_1 = 0, \; k = \frac{\Omega}{2}, \tag{11}$$

$$\psi_0 = -\frac{2}{A}\left(i\frac{\partial\psi_1}{\partial t} - \frac{\Delta}{2}\psi_1\right). \tag{12}$$

It is seen from (11) that here the solution is a superposition of two waves propagating in time towards each other:

$$\psi_1(t) = c_+ e^{ikt} + c_- e^{-ikt} \tag{13}$$

If there is an "incident wave" from the region $t < 0$, then after the "barrier" ($t > \tau$) we have (combining (12), (13) and boundary conditions, $\psi_0(0) = 1$, $\psi_1(0) = 0$):

$$\psi_0(\tau) = \cos(k\tau) + \frac{i\Delta}{\sqrt{\Delta^2 + A^2}}\sin(k\tau), \psi_1(\tau) = \frac{iA}{\sqrt{\Delta^2 + A^2}}\sin(k\tau). \tag{14}$$

Expression (14) implies the existence of an interference pattern for the populations of the basic states of a qubit under the effect of a unipolar pulse. The probability of pseudospin-flip ($|\downarrow\rangle \to |\uparrow\rangle$ transition) and the frequency of the Ramsey-type oscillations obey the expression (9). Expressions (9) and (14) also illustrate the fact that in order to obtain a contrasting picture of Ramsey interference, we need to meet the conditions of phase matching, which leads to the occurrence of standing waves over the time interval of the pulse action. Note that when someone discussing the phase synchronism in spatial problems of optics or quantum mechanics, such arguments allows understanding how the smoothing of parameter jumps at the boundaries affects the phase contrast. For example, based on such reasoning, it is possible to understand the influence of optical inhomogeneity of film boundaries on the width and position of resonances in the theory of the Fabry-Perot resonator.

In our case, the interpretation used allowed us to justify a fairly important conclusion: if the rise/fall times for the pulse is much less than its duration, then the contribution of switching on/off is small and can be taken into account by the perturbation theory.

A more complex character of a temporal interference pattern of qubit states population arises in the case of a pair of consecutive pulses. Here the pairs of waves arise both within time intervals of pulses action and in the gap between them, and their interference depends on the amplitudes and durations of the pulses.

Let us again begin our consideration with the case of the Rabi technique, so that the two "Rabi pulses" both with duration $\tau$ act on the system successively. And let the gap between them correspond to the time $\tau_R$. Free precession of the qubit during this gap is described by the expression

$$U_f = \begin{pmatrix} \exp(-i\omega\tau_R/2) & 0 \\ 0 & \exp(i\omega\tau_R/2) \end{pmatrix}. \quad (15)$$

The population of the excited state of a qubit after two successive pulses includes an interference term, which depends on the phase difference arising due to the delay of the second pulse. The evolution operator $U_{2f1} = U_2 U_f U_1$ here is as follows:

$$U_2 = e^{-iM_2}, M_2 = M_{2x}\sigma_x + M_{2y}\sigma_y + M_{2z}\sigma_z, M_{2x} = -A\tau\cos\varphi/4,$$
$$M_{2y} = -A\tau\sin\varphi/4, M_{2z} = -(\Delta - \omega)\tau/2, \varphi = \omega(\tau + \tau_R).$$

We can obtain the expression that was previously obtained by Ramsey [26] for pseudospin-flip probability by multiplication of the matrices. In the vicinity of resonance, where $A \gg \Delta - \omega$ and so approximate equality $A \approx \Omega_R$ is satisfied, the transition probability has the simple form:

$$W_{|0\rangle \to |1\rangle} = \frac{1}{2}\sin^2(\Omega_R\tau)\left(1 + \cos((\Delta - \omega)\tau_R)\right). \quad (16)$$

Therefore the maximum population of the excited state is achieved at $\Omega_R\tau = \pi/2$, and the interference pattern is determined by the parameter $(\Delta - \omega)\tau_R$.

For two successive unipolar pulses we obtain by analogy:

$$W_{|0\rangle \to |1\rangle} = 4\frac{A^2}{\Omega^2}\sin^2\left(\frac{\Omega\tau}{2}\right)\left(\cos\left(\frac{\Omega\tau}{2}\right)\cos\left(\frac{\Delta\tau_R}{2}\right) - \frac{\Delta}{\Omega}\sin\left(\frac{\Delta\tau_R}{2}\right)\sin\left(\frac{\Omega\tau}{2}\right)\right)^2. \quad (17)$$

It is seen that the pattern of Ramsey interference for a qubit under the action of a pair of unipolar pulses resembles the case of the action of two Rabi pulses considered above.

### C. Numerical simulations of a state control procedure of a single-qubit system

In the contour graph presented in Fig. 1, the color shows the population of the ground state of a single flux qubit, $W_0(t, A) = |\langle 0|\psi(t)\rangle|^2$, as a function of the amplitude and time from the onset of the action of a unipolar pulse. The results are obtained from numerical calculations of the equation (1). The parameters of the flux qubit are close to the works [3, 27]. The red regions correspond to the system in the ground state, and the violet regions correspond to the excited state. With an increase in the amplitude of the pulse, the frequency of the population oscillations increases, similar to what is observed using the Rabi technique. This result agrees well with the expression (9). In particular, trajectories on the plane of parameters $(A, t)$ that describe the complete pseudospin-flip of the system can be found. The black curves in Fig. 1 correspond to the dependencies $A = (\pi(1 + 2n))/2\tau$, $A \gg \Delta$, where $n$ is an integer. It's seen that if we can prepare a picosecond unipolar pulse with arbitrary amplitude and duration, then we can obtain any qubit state in a corresponding timescale.

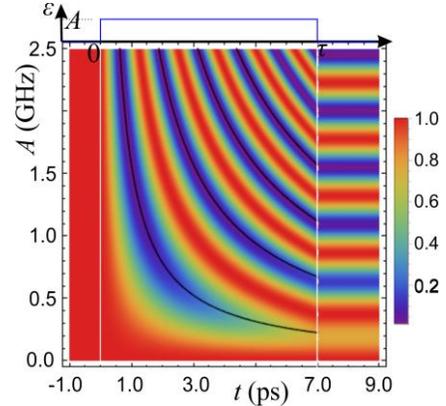

FIG. 1. The probability $W_0$ of the ground state population for a single qubit ($\Delta/h = 0.25$ GHz) as a function of the amplitude, $A$, and time $t$ from the onset of the action of a unipolar pulse. The black curves correspond to trajectories on the plane of parameters $(A, t)$ were the system acquires the complete pseudospin-flip. The white vertical lines indicate the beginning and the end of a unipolar pulse. The color scale of the population probability $W_0$ is given on the right side of the figure.

Next, we consider Ramsey interference under the impact of a pair of unipolar pulses with durations $\tau_1$ and $\tau_2$. For simplicity, we assume that both pulses have the same amplitude, $A$. Ramsey oscillations manifest themselves as periodic changes in the population of qubit states affected by the time delay $\tau_R$ between the two pulses (alternation of red and violet regions highlighted by a white box in the contour plot shown in Fig. 2(a)). The Bloch sphere shown in Fig. 2(b) illustrates the action of the pulses. The first pulse creates a superposition between the ground and the excited states (analogous to the $\pi/2$ pulse in Rabi technique for $A \gg \Delta$). This corresponds to the rotation of the state vector around the $x$ axis of the Bloch sphere (black curve, $P_1$, in Fig. 2(b)). Then the system freely evolves, which corresponds to the rotation of the state vector along the equator (blue curve, $F_E$, in Fig. 2(b)). The angular velocity of rotation is determined by the energy gap between the ground and excited states, $\Delta$. The second pulse transfers the qubit toward either the ground or excited state depending on the delay between the pulses, $\tau_R$ (green curve, $P_2$, in Fig. 2(b)). The calculated dependence of the ground state population versus the delay between pulses, $W_0(\tau_R)$, see Fig. 2(c) is in good agreement with the predictions of the obtained expression (17). It is seen that an arbitrary qubit state can be prepared also using two unipolar pulses of picosecond duration, rather than one. But in this case, pulses of smaller amplitude ($A < 0.5$ GHz for $\Delta = 0.25$ GHz) can be used, see white box at Fig. 2(a), that is important for the practical implementations.

### D. Numerical simulation of relaxation effects

There are several noise sources that are very important for the operation of quantum registers. First, it is noise caused by devices for initializing and readout of qubit states. Secondly, noise caused by the influence of the environment. In the context of this article, it is important for us that the decoherence does not destroy the Ramsey interference, so we will assume that the qubit was originally prepared in a pure state, for example, by a special cooling method, and the readout occurs with the minimum possible error by using a perfect resonator.

It is well known that charge fluctuations at Josephson contacts, flux fluctuations in a superconducting circuit, and radiation damping can be considered as the influence of a boson bath that produces phase and energy relaxation in a two-dimensional Hilbert subspace of qubit states. In this case, the equation for the density operator of the qubit $\rho$ in the Born-Markov approximation takes the following form [28, 29]:

$$\frac{\partial \rho}{\partial t} = i[\rho, H(t)] + \gamma\left(\sigma_-\rho\sigma_+ - \frac{1}{2}\{\sigma_+\sigma_-, \rho\}\right) + \gamma_\phi(\sigma_z\rho\sigma_z - \rho),$$

where $\gamma_\phi$ is the rate of the phase damping and the parameter $\gamma$ is responsible for the rate of energy loss, $\sigma_+ = \sigma_-^\dagger = |1\rangle\langle 0|$ are raising and lowering operators of the qubit. The transverse dephasing usually dominates over the energy relaxation $\gamma_\phi >> \gamma$ and we neglect temperature effects [30, 31].

According to [31], the flux qubits are characterized by phase dumping and relaxation times of ~ 100 μs and it is known that an increase in the influence of quantum noise leads to a broadening and overlapping of resonances. However, we note that the position of Ramsey fringes and the interference effects shown in Fig. 2 (d) are well observed even when taking into account the influence of the environment. Therefore using unipolar picosecond pulses we can measure for example the decoherence times with good accuracy based on Ramsay oscillations.

## III. TWO-QUBIT OPERATIONS

### A. Interference of states in two-qubit system

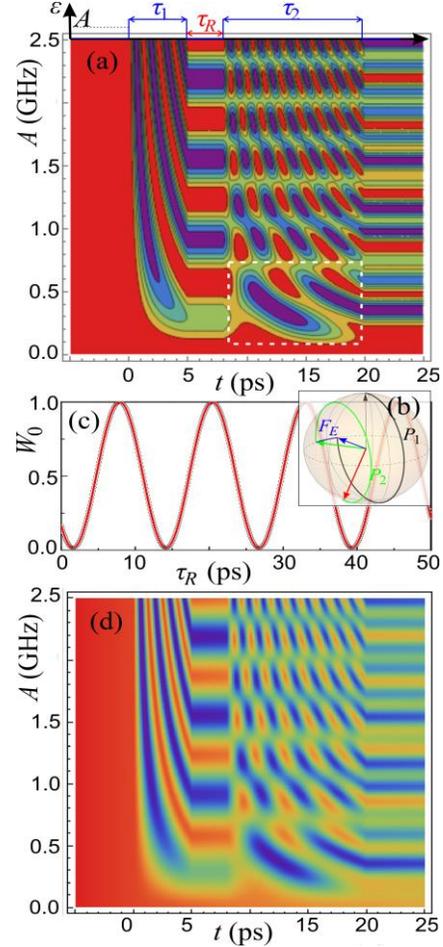

FIG. 2: (a) The color shows the probability of the ground state population of the qubit as a function of the amplitude $A$ and time $t$ under the action of a pair of unipolar pulses, $\tau_1 = 5$ ps, $\tau_2 = 12$ ps, $\tau_R = 3$ ps, $\Delta/h = 0.25$ GHz. (b) Image of the qubit state evolution on a Bloch sphere. (c) The calculated analytically by Eq. (17) (red curve) and numerically (black circles) dependence of the ground state population versus the delay between pulses. (d) However, we note that the position of Ramsey fringes and the interference effects, shown in Fig. 2 (d), are well defined for $\gamma_\phi = 0.1$ GHz and $\gamma = 0.05$ GHz.

Let us consider a simple register consisting of two interacting flux qubits with the following Hamiltonian:

$$H(t) = H^{(1)}(t) \otimes I^{(2)} + I^{(1)} \otimes H^{(2)}(t) - \frac{1}{2}J(t)\sigma_x^{(1)} \otimes \sigma_x^{(2)}, \quad (18)$$

where $H^{(i)}$ is the Hamiltonian of the $i$-th qubit ($i = 1, 2$), see Eq. (3), and $J(t)$ is a factor characterizing the interaction of qubits. This Hamiltonian (18) can be written in matrix form as follows:

$$H(t) = -\frac{1}{2}\begin{pmatrix} \Delta^{(1)}+\Delta^{(2)} & \varepsilon^{(2)}(t) & \varepsilon^{(1)}(t) & J(t) \\ \varepsilon^{(2)}(t) & \Delta^{(1)}-\Delta^{(2)} & J(t) & \varepsilon^{(1)}(t) \\ \varepsilon^{(1)}(t) & J(t) & -\Delta^{(1)}+\Delta^{(2)} & \varepsilon^{(2)}(t) \\ J(t) & \varepsilon^{(1)}(t) & \varepsilon^{(2)}(t) & -\Delta^{(1)}-\Delta^{(2)} \end{pmatrix}. \quad (19)$$

We assume that the register is affected by rectangular unipolar pulses. The state of the system can be presented as:

$$|\psi(t)\rangle = \psi_1(t)|\downarrow\downarrow\rangle + \psi_2(t)|\uparrow\downarrow\rangle + \psi_3(t)|\downarrow\uparrow\rangle + \psi_4(t)|\uparrow\uparrow\rangle,$$

where $|\psi_1(t)|^2 + |\psi_2(t)|^2 + |\psi_3(t)|^2 + |\psi_4(t)|^2 = 1$.

In the frame of the approach presented in the previous section, the domain of the four-component wave function, $\psi_j(t)$ ($j = 1, 2, 3, 4$), can be divided into three parts, ($t < 0$, $0 < t < \tau$ and $t > \tau$), with three matrix Schrödinger equations with constant coefficients. One can define the eigenvalues, $\lambda_j$, and eigenvectors, $v_j$, of the matrix (19). Each component of the wave function will be a superposition of four waves propagating backward and forward in time:

$$\psi_j(t) = \sum_{j=1}^{4} c_{jk} v_k e^{-i\lambda_k t}. \quad (20)$$

The expansion coefficients, $c_{jk}$, are again determined by matching the waves at the edges of the pulse. The population of each normal mode can be represented as a Ramsey interference pattern. Solution (20) allows us to write an expression for the evolution operator of a two-qubit system under the effect of a unipolar pulse in the form:

$$U(t) = \sum_{k=1}^{4} |v_k\rangle e^{-i\lambda_k t} \langle v_k|, \ 0 < t < \tau. \quad (21)$$

Again, to get a contrasting picture of Ramsey interference, we need to meet the conditions for phase matching. We show that for a series of rectangular pulses, these conditions can be met for four waves.

Let us consider an illustrative particular case that allows an explicit solution. In fact, we can treat this case as the implementation of the simplest two-qubit operation, which is induced by a special coupler. Let the qubits be identical ($\Delta^{(1)} = \Delta^{(2)} \equiv \Delta$), "non-disturbed" ($\varepsilon^{(1)} = \varepsilon^{(2)} = 0$), and the control pulse only turns on the coupling between them (*Example 1*):

$$H = -\frac{1}{2}\begin{pmatrix} 2\Delta & 0 & 0 & J \\ 0 & 0 & J & 0 \\ 0 & J & 0 & 0 \\ J & 0 & 0 & -2\Delta \end{pmatrix}, \ 0 < t < \tau. \quad (22)$$

The eigenvalues / eigenvectors of Hamiltonian (22) are:

$\lambda_1 = -J$, $\lambda_2 = J$, $\lambda_3 = \sqrt{J^2 + 4\Delta^2}$, $\lambda_4 = -\sqrt{J^2 + 4\Delta^2}$, $v_1^T = (0, -1, 1, 0)$, $v_2^T = (0, 1, 1, 0)$, $v_3^T = \left(-\left(2\Delta + \sqrt{J^2 + 4\Delta^2}\right)/J, 0, 0, 1\right)$, $v_4^T = \left(-\left(2\Delta - \sqrt{J^2 + 4\Delta^2}\right)/J, 0, 0, 1\right)$. Let the initial state of the register be: $|\downarrow\downarrow\rangle$, with energy equal to $-\Delta$. Transition

probability to the state $|\uparrow\uparrow\rangle$ (with energy equal to $+\Delta$) is as follows:

$$W_{|\downarrow\downarrow\rangle \to |\uparrow\uparrow\rangle} = \frac{J^2}{J^2 + 4\Delta^2} \sin^2\left(\tau\sqrt{J^2 + 4\Delta^2}\right). \quad (23)$$

It is seen that the populations of the selected levels oscillate with the frequency $\sqrt{J^2 + 4\Delta^2}$. A complete spin-flip occurs at the time $\tau \approx \pi/2J$ ($J \gg \Delta$).

It is also interesting to consider interference effects arising from phase mixing as a result of the action of several unipolar pulses, by analogy with a single-qubit system (*Example 2*). Let the qubits again be identical, "non-disturbed", and the first control pulse only turns on the coupling between them:

$$U_1 = \begin{pmatrix} 1 & 0 & 0 & -ig \\ 0 & 1 & -ig & 0 \\ 0 & -ig & 1 & 0 \\ -ig & 0 & 0 & 1 \end{pmatrix}, \quad \begin{array}{l} J \gg \Delta, \\ J\tau_1 \ll 1, \\ g = \frac{J\tau_1}{2}. \end{array} \quad (24)$$

After this ($J(t > \tau_1) = 0$), the qubits of the register are simultaneously affected by unipolar pulses with equal durations, $\tau_2$, but different amplitudes. The eigenvalues and eigenvectors of the Hamiltonian (19) can be found as follows:

$$\tilde{\lambda}_1 = -\frac{|E_1 - E_2|}{2}, \ \tilde{\lambda}_2 = \frac{|E_1 - E_2|}{2}, \ \tilde{\lambda}_3 = -\frac{|E_1 + E_2|}{2}, \ \tilde{\lambda}_4 = \frac{|E_1 + E_2|}{2},$$

$$\tilde{v}_1^T = \left(\frac{-E_1 E_2 + \Delta^2 - \Delta|E_1 - E_2|}{\varepsilon^{(1)}\varepsilon^{(2)}}, \frac{\Delta - E_2\text{sign}[E_1 - E_2]}{\varepsilon^{(1)}}, \frac{\Delta + E_2\text{sign}[E_1 - E_2]}{\varepsilon^{(1)}}, 1\right),$$

$$\tilde{v}_2^T = \left(\frac{-E_1 E_2 + \Delta^2 - \Delta|E_1 - E_2|}{\varepsilon^{(1)}\varepsilon^{(2)}}, \frac{\Delta + E_2\text{sign}[E_1 - E_2]}{\varepsilon^{(1)}}, \frac{\Delta - E_2\text{sign}[E_1 - E_2]}{\varepsilon^{(1)}}, 1\right),$$

$$\tilde{v}_3^T = \left(\frac{E_1 E_2 + \Delta^2 - \Delta|E_1 - E_2|}{\varepsilon^{(1)}\varepsilon^{(2)}}, \frac{\Delta - E_2\text{sign}[E_1 + E_2]}{\varepsilon^{(1)}}, \frac{\Delta - E_2\text{sign}[E_1 + E_2]}{\varepsilon^{(1)}}, 1\right),$$

$$\tilde{v}_4^T = \left(\frac{E_1 E_2 + \Delta^2 + \Delta|E_1 - E_2|}{\varepsilon^{(1)}\varepsilon^{(2)}}, \frac{\Delta + E_2\text{sign}[E_1 + E_2]}{\varepsilon^{(1)}}, \frac{\Delta + E_2\text{sign}[E_1 + E_2]}{\varepsilon^{(1)}}, 1\right),$$

where $E_{1,2} = \sqrt{A_{1,2}^2 + \Delta^2}$. Next, we can write the evolution operator for the register during the pulses impact (normalizing the eigenvectors written above):

$$U_2 = \sum_{k=1}^{4} |\tilde{v}_k\rangle e^{-i\tilde{\lambda}_k \tau_2} \langle \tilde{v}_k|. \quad (25)$$

After the end of the pulses, the control short pulse of large amplitude is again applied to the coupler, $U_3 = U_1$. Thus, in this case

$$W_{|\downarrow\downarrow\rangle \to |\uparrow\uparrow\rangle} = \frac{1}{4E_1^2 E_2^2}\left(D_1 \cos^2\left[\frac{E_1\tau_2}{2}\right]\cos^2\left[\frac{E_2\tau_2}{2}\right] + \right.$$
$$\left. D_2 \sin^2\left[\frac{E_1\tau_2}{2}\right]\sin^2\left[\frac{E_2\tau_2}{2}\right] - D_3 \sin[E_1\tau_2]\sin[E_2\tau_2]\right)^2, \quad (26)$$

where $D_1 = 4E_1^2 E_2^2 J^2 \tau_1^2$, $D_2 = 4J^2\tau_1^2\Delta^4 + (-2 + J^2\tau_1^2)^2 A_1^2 A_2^2$, and $D_3 = 2E_1 E_2 J^2 \tau_1^2 \Delta^2$. If the frequencies of the qubits are small with respect to the amplitudes of unipolar pulses, then

$$W_{|\downarrow\downarrow\rangle \to |\uparrow\uparrow\rangle} \simeq J^2\tau_1^2 \cos^2\left[\frac{E_1\tau_2}{2}\right]\cos^2\left[\frac{E_2\tau_2}{2}\right] + \frac{1}{4}\left(-2 + J^2\tau_1^2\right)^2 \sin^2\left[\frac{E_1\tau_2}{2}\right]\sin^2\left[\frac{E_2\tau_2}{2}\right]. \quad (27)$$

## B. Numerical simulations of a state control procedure of a two-qubit system

We begin the discussion of the numerical simulations of fast operations with the register from *Example 1* considered in the previous subsection: $\varepsilon^{(i)}(t) = 0$ and $J(t) = J\Theta(t)\Theta(\tau - t)$. The obtained dependence for the transition probability from the ground to the highest excited state versus the pulse duration acting only on the coupler is in good agreement with the analytical expression (23), see Fig. 3. In this case the black curves correspond to the maxima probability where $J = \pi(1 + 2n)/2\tau$, $J \gg \Delta$ ($n$ is an integer). We perform the states in the register and the inversion operation $|\downarrow\downarrow\rangle \to |\uparrow\uparrow\rangle$ with high accuracy (up to 99.9 %) by controlling only the coupling strength between qubits.

In *Example 2* (the unipolar pulse of duration $\tau_1$ is used to turn on the interaction between qubits, after which we simultaneously apply unipolar pulses of equal durations $\tau_2$ to both qubits, and then we turn on the interaction again) the dynamics of the entangled states substantially depends on the amplitudes of the pulses applied to qubits, see Fig. 4. The positions of the maxima (minima) of the Ramsey interference pattern are determined by the analytical expression (27).

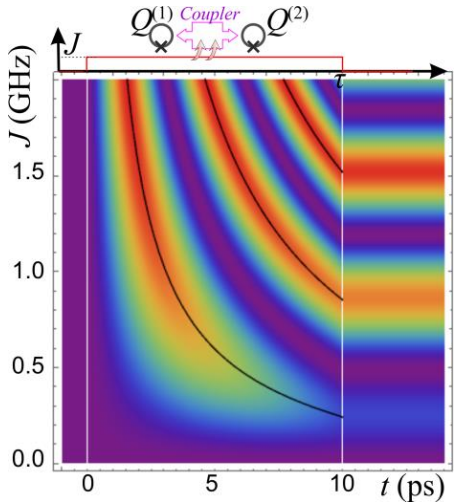

FIG. 3. The color shows the probability of the transition $|\downarrow\downarrow\rangle \to |\uparrow\uparrow\rangle$ in the register. One unipolar pulse acts on the coupler, $\tau = 10$ ps, see white vertical lines. The qubit parameters are as follows: $\Delta^{(i)}/h = 0.1$ GHz, $\varepsilon^{(i)} = 0$. The color scale is shown in Fig. 1.

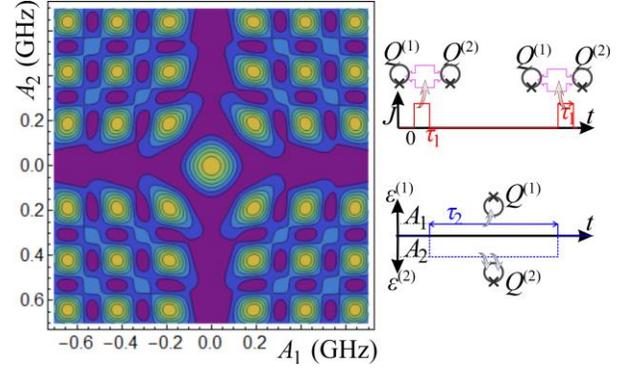

FIG. 4. The color shows the probability of the transition $|\downarrow\downarrow\rangle \to |\uparrow\uparrow\rangle$ in the register. Unipolar pulse is used to turn on the interaction ($0 \le t < \tau_1$, $\tau_1 = 1$ ps, $J/h = 1$ GHz), after which we simultaneously apply unipolar pulses $\varepsilon^{(i)}(t)$ of duration $\tau_2 = 30$ ps to both qubits $\Delta^{(i)}/h = 0.1$ GHz ($\tau_1 \le t \le \tau_2 + \tau_1$, $J = 0$), and then we turn on the interaction again for the time $\tau_1$ ($\tau_1 + \tau_2 < t \le 2\tau_1 + \tau_2$, $J/h = 1$ GHz). The color scale is shown in Fig. 1.

Further, we investigate numerically a case of initialization of the state of the register with constant coupling between qubits, and with a pair of unipolar pulses acting on the system. A similar situation is typical for digital-analog quantum computing [32-34]. For simplicity, let the first pulse acts on the first qubit in the register and the second pulse acts on the second one. We fixed the duration of the first unipolar pulse, $\tau_1$, and the time delay $\tau_R$. Changes in the population of levels for various amplitude of pulses ($A_1 = A_2$) and for a chosen durations of the second pulse, $\tau_2$, are shown in Fig. 5. The accuracy of the inversion $|\downarrow\downarrow\rangle \to |\uparrow\uparrow\rangle$ (Fig. 5(d,h)) and the read $|\downarrow\downarrow\rangle \to |\downarrow\downarrow\rangle$ (Fig. 5(a,e)) operations is 99.9 %, while for the transitions to $|\downarrow\downarrow\rangle \to |\uparrow\downarrow\rangle$ (Fig. 5(b,f)) and $|\downarrow\downarrow\rangle \to |\downarrow\uparrow\rangle$ (Fig. 5(c,g)), it is 99.5-99.9 %. The regions of the parameters of unipolar pulses required for excitation of various states (shown by white lines $\varepsilon^{(i)}(t)$ in Fig. 5(a-d)) do not intersect on the plane $(A_2, t)$ which allows controlling the register with high selectivity.

## IV. IMPLEMENTATION OF CONTROL PULSES WITH PICOSECOND DURATION

This section presents a possible engineering solution for implementation of picosecond unipolar magnetic flux pulses close to a rectangular shape with adjustable duration and amplitude. The control pulse is shaped from SFQ pulse, which can be generated, for example, using a DC-to-

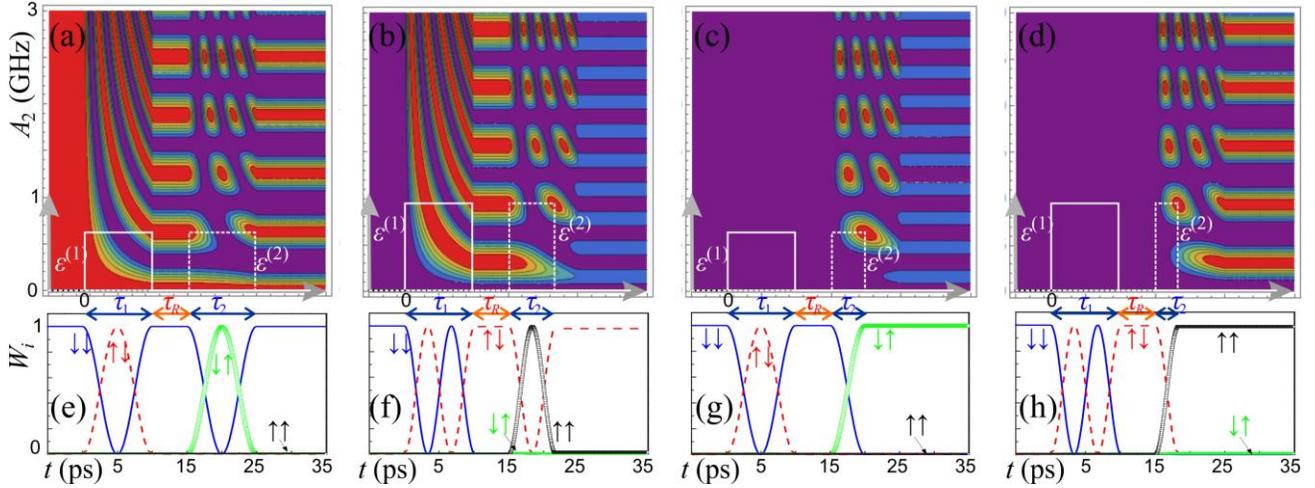

FIG. 5. The colors for panels (a–d) (and the marked curves for panels (e–h)) represents the population of the register levels under the action of a pair of pulses. Transitions $|\downarrow\downarrow\rangle \rightarrow |\downarrow\downarrow\rangle$ – (a) and (e), $|\downarrow\downarrow\rangle \rightarrow |\uparrow\downarrow\rangle$ – (b) and (f), $|\downarrow\downarrow\rangle \rightarrow |\downarrow\uparrow\rangle$ – (c) and (g), $|\downarrow\downarrow\rangle \rightarrow |\uparrow\uparrow\rangle$ – (d) and (h). $\Delta^{(1)}/h = 0.1$ GHz, $\Delta^{(2)}/h = 0.12$ GHz, $J/h = 0.01$ GHz, $\tau_1 = 10$ ps, $\tau_R = 5$ ps. The amplitude of pulses and the duration of the second pulse are as follows: $A_1/h = A_2/h = 0.63$ GHz (e, g), and $\tau_2 = 10$ ps (e) and 5 ps (g); $A_1/h = A_2/h = 0.93$ GHz (f, h), and $\tau_2 = 6.5$ ps (f) and 3.2 ps (h). The color scale is shown in Fig. 1.

SFQ converter [35]. The shaping is done by means of a special coupler, shown in Fig. 6(a).

Here SFQ pulse passes through a long Josephson junction (LJJ) having critical current $I_c$ and damping coefficient $\alpha = \omega_p/\omega_c = 0.05$ (where $\omega_p$ is the plasma frequency and $\omega_c$ is the characteristic frequency of LJJ). The edges of one of LJJ's electrodes are closed into a coupling loop. The length of the LJJ noticeably exceeds the size of the Josephson vortex so that the magnetic flux in the coupling loop is constant during SFQ passage through the coupler, and abruptly changes at the moments of SFQ entry and exit. The details of numerical simulations of SFQ propagation along LJJ can be found in our previous works [20, 21, 36]. We calculate the duration of the magnetic flux pulse adjusted by LJJ bias current, $i_b = I_b/I_c$, using the same method as in [20, 21], see Fig. 6(b).

The amplitude of the resulted magnetic flux pulse is adjusted by its transmission through a symmetrical superconducting circuit composed of two single-junction superconducting interferometers [37, 38]. These interferometers are identical except Josephson junctions (JJs). One JJ is a conventional one having critical current $I_{C0}$, while another is magnetic Josephson junction (MJJ) with a controllable critical current $I_{c1} = I_{cM}/I_{c0}$ [39-41]. The value of the damping coefficient for both junctions is $\alpha = 3$. The common inductance of the interferometers is magnetically coupled to a qubit, and is connected to the point of symmetry of the circuit. Therefore, if the critical currents of the JJs are equal, there is no output current in this inductance, and no flux transfer to the qubit, correspondingly. However, when the critical currents values deviate from each other, the output current is not zero, and its magnitude is proportional to their difference, see Fig. 6(c).

The proposed coupler allows an independent setting of both parameters of the output magnetic flux pulse with sharp fronts over fairly wide ranges. In the absence of control pulses, the qubit is not affected by stray currents in the control circuit.

We calculate the effect of output magnetic flux pulses on a two-qubit system to verify the validness of the proposed circuit. The durations and amplitudes of the pulses are fitted to create the necessary population of the levels. Fig. 6(d) illustrates a possibility of the inversion of a two-qubit system with a fidelity of >99 % where each qubit is affected by two sequential pulses with a delay of $\tau_R = 30$ ps. The formation of a entanglement state $|\downarrow\downarrow\rangle \rightarrow (|\downarrow\uparrow\rangle + |\uparrow\uparrow\rangle)/\sqrt{2}$ achieved with a similar technique is shown in Fig. 6(e).

The characteristic operation time depends on the plasma frequency of the coupler Josephson junctions, $\omega_p$, which in turn is proportional to the root of their critical current density. Superconducting technology allows the implementation of Josephson junctions for various applications in the frame of standard fabrication processes with JJ critical current densities from 30 A/cm² to 20 kA/cm². Consequently, the plasma frequency lies in the range from ~ 25 GHz to 465 GHz, respectively.

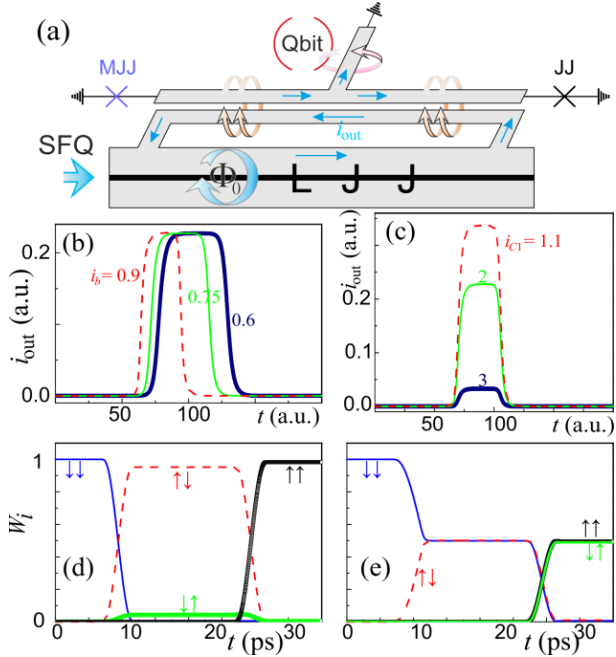

FIG. 6. (a) Sketch of the coupler converting SFQ into rectangular magnetic flux pulse with independently adjustable duration and amplitude. (b) LJJ bias current, $i_b$, effect on the output current pulse duration. (c) Effect of MJJ critical current, $I_{c1}$, variation on the output current pulse amplitude. The probabilities of the population of two-qubit system levels under excitation by magnetic flux pulses formed by proposed coupler: (d) the inversion, (e) the formation of an entangled state with the utilization of the presented coupler. Currents are normalized to conventional JJ critical current but LJJ bias current, $i_b$, is normalized to LJJ critical current. The damping coefficient is $\alpha = 0.05$ for LJJ and $\alpha = 3$ for JJ and MJJ.

The width of the fluxon is about 4 Josephson lengths. For the parameters of the control cell elements (bias current, damping coefficient, etc.) used in the simulation, SFQ velocity in LJJ is half of the maximum possible (Swihart) velocity. To estimate the duration of the control pulse front, one has to divide the fluxon width by its speed, and to convert to dimensional units, divide the result by the cyclic plasma frequency. For the mentioned critical current densities, the front width is from 50 to 3 ps. Assuming that the pulse width should be an order larger than the width of its front, one can obtain the operation duration of just a few tens of picoseconds.

While the critical current of LJJ is assumed to be large (several mA), the effect of thermal noise current of nA scale, typical for mK temperature, on the fluxon propagation and so, as well, on the pulse shape is assumed to be negligible. The vanishingly small effect is expected also from quantum noise considering fluxon energy and the presented time of the process. Since the accuracy of the bias current setting lies in the same nA scale for standard equipment, we conclude that the considered sources of noise can hardly affect the relatively high-energy fluxon.

Practical time limitation in the considered control scheme is expected to arise from time jitter of switching of Josephson junctions in digital cells which generate and route control SFQ pulses to the qubits (like SFQ splitter, JTL cells and SFQ drivers providing matching of impedance of digital cells with passive transmission lines). In accordance with the literature [42,43] and our previous study [44] one can expect sub-ps time precision of the driving circuit.

Thus, the proposed schematic provides a robust solution for manipulation of the qubits states at picosecond timescale.

## V. CONCLUSION

In conclusion, we proposed an ultrafast qubit control concept allowing manipulations with a single and multiple superconducting qubits using short unipolar rectangular pulses with durations much less than inverse qubit transition frequency. These pulses create an interference pattern for populations of qubit levels, which is completely equivalent to the formation of Ramsey fringes obtained by using the modulated high-frequency pulses in the Rabi technique. Ramsey interference can lead to the formation of a sharp phase contrast for populations in a multi-level system. Physically, this effect is caused by the interference of waves in time, that we have justified on the basis of analytical consideration and numerical modeling. This fact allows us to propose a protocol utilizing two successive unipolar pulses for the initialization of arbitrary single-qubit and two-qubit register states. Pulses with sharp fronts can create a perfect contrast of populations. Transient regions of short control duration have little effect on the interference pattern. We proposed a notional engineering solution for transformation of SFQ pulses into the required unipolar pulses with sharp fronts and independently adjustable amplitude and duration. This makes it possible to control the quantum states of the considered systems with the fidelity of more than 99% on a picosecond timescale. The straightforward consequence of the application of the proposed concept could be a dramatic reduction of time of single- and multiple-qubit gates which is favorable for the implementation of error correction in quantum circuits using dynamical decoupling [45].

## ACKNOWLEDGMENTS

This work was supported by the RFBR grant No. 20-07-00952. Section 4 is prepared with the support of RFBR grant No. 19-32-90208. The literature review is supported by the grant of the President of the Russian Federation (MD-186.2020.8). V.R. acknowledges the Basis Foundation scholarship.

* E-mail: igor.soloviev@gmail.com